\documentclass[11pt]{article}

\usepackage{graphicx}

\begin{document}

\title{\bf Climbing Mount Scalable: \\
Physical-Resource Requirements \\
for a Scalable Quantum Computer}

\author{
Robin Blume-Kohout,$^{\hbox{\small (1)}}$
Carlton M.~Caves,$^{\hbox{\small (2)}}$
and Ivan H.~Deutsch$^{\hbox{\small (2)}}$
\\ \hbox{} \\
$^{\hbox{\small (1)}}$\em Los Alamos National Laboratory, Mail Stop B210, \\
\em Los Alamos, NM~87545, USA \\
$^{\hbox{\small (2)}}$\em Department of Physics and Astronomy,
University of New Mexico, \\
\em Albuquerque, NM~87131--1156, USA
\\ \hbox{} \\
\em E-mail: caves@info.phys.unm.edu, ideutsch@info.phys.unm.edu
}

\date{September 25, 2002}

\maketitle

\vspace{36pt}
\noindent
The primary resource for quantum computation is Hilbert-space
dimension.  Whereas Hilbert space itself is an abstract construction,
the number of dimensions available to a system is a physical quantity
that requires physical resources.  Avoiding a demand for an
exponential amount of these resources places a fundamental constraint
on the systems that are suitable for scalable quantum computation. To
be scalable, the effective number of degrees of freedom in the
computer must grow nearly linearly with the number of qubits in an
equivalent qubit-based quantum computer.

\vspace{12pt}
\vspace{24pt}

\section{INTRODUCTION}
\label{sec:intro}

Quantum computation is an alluring long-term goal for the emerging
field of quantum information science [1].  In this paper we address
the question of what physical resources are required for quantum
computation and, in particular, how the required resources scale with
problem size.  We explicitly do {\em not\/} try to establish whether
quantum computing is more powerful than classical computing, that
being an unsolved problem in computational complexity theory.  Rather
we {\em assume\/} that quantum computing is more powerful than
classical computing, and given this assumption, we ask how the
physical resources required to take advantage of the power of quantum
computing scale with problem size.  By determining how to avoid a
physical-resource demand that increases exponentially with problem
size, we establish necessary conditions for a physical system to be a
scalable quantum computer.

The initial step in a quantum computation [2] is to store classical
information (the input) as some quantum state of the computer.  The
computer then runs through a carefully controlled sequence of unitary
operations and/or measurements (the program).  A program can be
carried out wholly by reversible unitary operations [3] or wholly by
irreversible quantum measurements [4--6].  Generally both will be
used, especially in implementing quantum error correction [7--9] and
fault-tolerant quantum computation [10--13].  At the completion of
the computation, the answer (the output) is stored as classical
information that can be read out with high probability by a
measurement.  The power of a quantum computer lies somewhere in the
murky region between the classical input and the classical output---a
region where classical, realistic descriptions fail.

Ask for the crucial property of that murky region, and you will get
nearly as many answers as there are quantum information scientists:
the superposition principle of quantum mechanics and associated
quantum interference and quantum parallelism; quantum entanglement;
the use of entangling unitary operations; the collapse of the wave
function after measurement and associated information-disturbance
trade-offs.  All of these distinguish quantum systems from classical
ones.  How are we to decide which is the crucial quantum feature?

There are quantum-information-processing scenarios for which one or
more of these features can be identified up front as a key resource
not available in the comparable classical situation.  Examples
include quantum cryptography [14] and quantum communication
complexity or distributed computing [15], where there are clearly
identified, separate parties who can do things locally, but who
interact with one another only through exchange of classical
information.  In quantum key distribution, for example, two parties
seek to generate a secret key that can be used for secure
communication.  The presence of an eavesdropper is revealed by the
disturbance produced when she obtains information about the key.  In
quantum communication complexity or distributed computing, separate
parties try to perform some computational task through local
operations and classical communication.  Prior entanglement is a
resource not available classically, which allows the parties to do
things, such as teleportation, that can't be done classically.

A quantum computer is not like these examples, however, because there
are no clearly identified separate parties.  Although it is natural
to think in terms of a division of a quantum computer into parts that
exchange quantum information, these parts are not like separate
parties: they generally are not well separated, and they interact
quantum mechanically with one another.  Moreover, the division into
subsystems is not unique [16], and there are proposals for quantum
computing that have no natural division into parts at all [17--19].

Given that none of the quantum features listed above stands out as
the source of a quantum computer's power, we argue that the
empowerment stems from the murky region itself: a quantum computer
can escape the bounds of classical information processing because
there is no efficient realistic description of what happens between
the classical input and the classical output.  We do not know how to
characterize completely the circumstances for which there is no
efficient classical description, since knowing this would be
equivalent to knowing when a quantum computation is more powerful
than a classical computation.  If quantum computers are more
powerful, however, their ability to access arbitrary states in
Hilbert space leads to such situations.

It is difficult to pin down the source of a quantum computer's power
because arbitrary states can be accessed in very different physical
systems---different hardware---using very different control
techniques---different software, e.g., the use of reversible unitary
operations vs.\ irreversible quantum measurements.  Yet no matter how
a quantum computation is packaged, we can identify one universal
prerequisite: the computer must have a Hilbert space large enough to
accommodate the computations.  If the computer is to be a
general-purpose computer, able in principle to solve problems of
arbitrary size, it must have a Hilbert space whose dimension is
capable in principle of growing exponentially with problem size.
Hilbert space is essential for quantum computation, and the primary
resource is {\em Hilbert-space dimension}.

Hilbert spaces of the same dimension are {\em fungible}.  What can be
done in one can be done in principle in any other of the same
dimension: simply map one Hilbert space onto the other, including all
the subsystems, operations, and measurements.  Which Hilbert space is
used to represent and process quantum information only becomes
important when further physical considerations are introduced.  What
is important at the outset is that a system have ``quantum
information inside,'' i.e., that there be information stored as
arbitrary states in the system's Hilbert space.

Though Hilbert spaces are fungible, the physical systems described by
those Hilbert spaces are not, because {\em we don't live in Hilbert
space}, or as Asher Peres puts it [20], ``Quantum phenomena do not
occur in Hilbert space. They occur in a laboratory.''  A Hilbert
space gets its connection to the world we live in through the
physical quantities---position, linear momentum, energy, angular
momentum---of the system that is described by that Hilbert space.
These physical quantities arise naturally from spacetime symmetries
and the system Hamiltonian, and they are the physical resources that
must be supplied to access various parts of the system Hilbert space.
The crucial {\em physical\/} question for quantum computation is the
following: {\em how much of these resources is required to achieve a
Hilbert-space dimension sufficient for a computation?}  This is the
question we address in this paper.

Quantum mechanics---and its generalization to quantum
fields---constrains our description of physical systems sufficiently
that we can formulate the question of physical-resource demands in a
general way.  We find that to avoid supplying an amount of some
physical resource that grows exponentially with problem size, the
computer must be made up of parts---degrees of freedom in the
simplest analysis, particles and field modes acting as effective
degrees of freedom in the case of quantum fields---whose number grows
nearly linearly with the number of qubits required in an equivalent
quantum computer.  This thus becomes a fundamental requirement for a
system to be a {\em scalable\/} quantum computer [21].

This result will not be a surprise to researchers in quantum
information science.  Indeed, it is often assumed {\em a priori\/}
that a quantum computer must be made up of interacting parts.  Our
analysis here provides a general justification for this requirement,
based only on an examination of how Hilbert-space dimension is
related to physical resources in different physical systems.

We emphasize that this requirement is an initial barrier that must be
surmounted by proposals for scalable quantum computation, before such
proposals confront the difficult tasks of initialization, control,
protection from errors, and readout, to which we return at the end.
Surmounting this barrier does not guarantee that a proposal can meet
the further requirements; it is a necessary, but by no means
sufficient requirement for a scalable quantum computer.  An important
point of this paper is that one can draw general conclusions about
the physical systems that can be used for quantum computation just by
considering whether the system can efficiently provide the {\em
primary\/} resource of Hilbert-space dimension, without getting
enmeshed in questions about the other necessary requirements for the
operation of a quantum computer.

The remainder of the paper is organized as follows.  In
Sec.~\ref{sec:dofanalysis} we consider the physical resources
required by a quantum computer that is built out of subsystems that
are separate ``degrees of freedom.''  The conclusions drawn there
form the core of our analysis, which we extend in Sec.~\ref{sec:Fock}
to systems that require a more general description in terms of
quantum fields.  In Sec.~\ref{sec:otherreq} we consider how our
necessary condition for scalable physical resources is related to
other requirements for quantum computation, including initialization,
control, stability, and measurement.  These requirements touch on
some of the more difficult issues in trying to pinpoint the source of
a quantum computer's power, including the role of entanglement and
the scalability of quantum computers that use highly mixed states. In
Sec.~\ref{sec:conclusion}, we summarize our conclusions.

\section{DEGREES-OF-FREEDOM ANALYSIS OF RESOURCE \\ REQUIREMENTS}
\label{sec:dofanalysis}

\subsection{The Role of Planck's Constant}
\label{sec:Planck}

Dimensionless quantities in physics are determined by writing the
relevant physical quantities in terms of a relevant scale.  For the
dimension of a system's Hilbert space, Planck's constant $h$ sets the
scale; the available number of Hilbert-space dimensions is determined
by writing an appropriate combination of physical quantities, the
{\em action}, in units of $h$.

The analysis of resource demands is particularly simple for systems
of particles described by ordinary quantum mechanics, i.e., not
requiring the more general description in terms of quantum fields.
For these systems, the subsystems can be identified with the {\em
degrees of freedom\/} of the particles.  The quantum state of such a
computer is described in a Hilbert space that is a tensor product of
the Hilbert spaces of the degrees of freedom.

A degree of freedom corresponds to a pair of (generalized) canonical
co\"ordinates, position $q$ and momentum $p$.  The physical resources
are the ranges of positions and momenta, $\Delta q$ and $\Delta p$,
used by the computation.  The physically relevant measure of these
resources is the corresponding phase-space area or {\em action},
$A=\Delta q\Delta p$.  For a degree of freedom that is an intrinsic
angular momentum $J$, we can use $\Delta q=2\pi$ and $\Delta p=\Delta
J$, thus giving $A=2\pi\Delta J$.  The connection to Hilbert space
comes from the fact that a quantum state occupies an area in phase
space given by Planck's constant $h\,$; orthogonal states correspond
roughly to nonoverlapping areas, each of area $h$ [22].  Thus the
available dimension of the Hilbert space for a single degree of
freedom is given approximately by $A/h$.  The goal of scalability is
to avoid having to supply an action resource $A$ for any degree of
freedom that grows exponentially with problem size.

\subsection{Degrees-of-Freedom Analysis}
\label{sec:dof}

We measure the Hilbert-space dimension required for a quantum
computation in qubit units: let the problem size for a computation be
$n$, and let $N={\bf F}(n)$ be the number of qubits required for the
computation, assuming an optimal qubit algorithm that requires only a
polynomial number of qubits, an example being Shor's factoring
algorithm [2].  Here and throughout, bold type denotes a function
that is bounded above by a polynomial.  The Hilbert-space dimension
needed for the computation is $2^N=2^{\mbox{\scriptsize\bf F}(n)}$.
Using qubit units, we see that the Hilbert-space dimension grows
exponentially with problem size. We assume that there is no more
efficient algorithm in a Hilbert space with some other structure than
the qubit tensor-product structure, this being part of our assumption
that Hilbert spaces are fungible.

Suppose now that the $j$th degree of freedom supplies an action
$A_j$.  The Hilbert space of the entire system is a tensor product of
the Hilbert spaces for the degrees of freedom, so the overall Hilbert
space has dimension
\begin{equation}
2^N\sim{A_1\over h}\cdots{A_T\over h}={V\over h^T}\;,
\end{equation}
where $V$ is the phase-space volume used by the computation. If $T$
grows more slowly than linearly with $N$ (within specific logarithmic
corrections discussed below), at least one of the actions must grow
exponentially with $N$, thus requiring an exponential amount of some
physical resource.  In contrast, if $T$ grows linearly with $N$, no
degree of freedom has to supply an increasing amount of action, which
makes the system a candidate for a scalable quantum computer.

\begin{figure}[p]
\begin{center}
\includegraphics[height=3.5in]{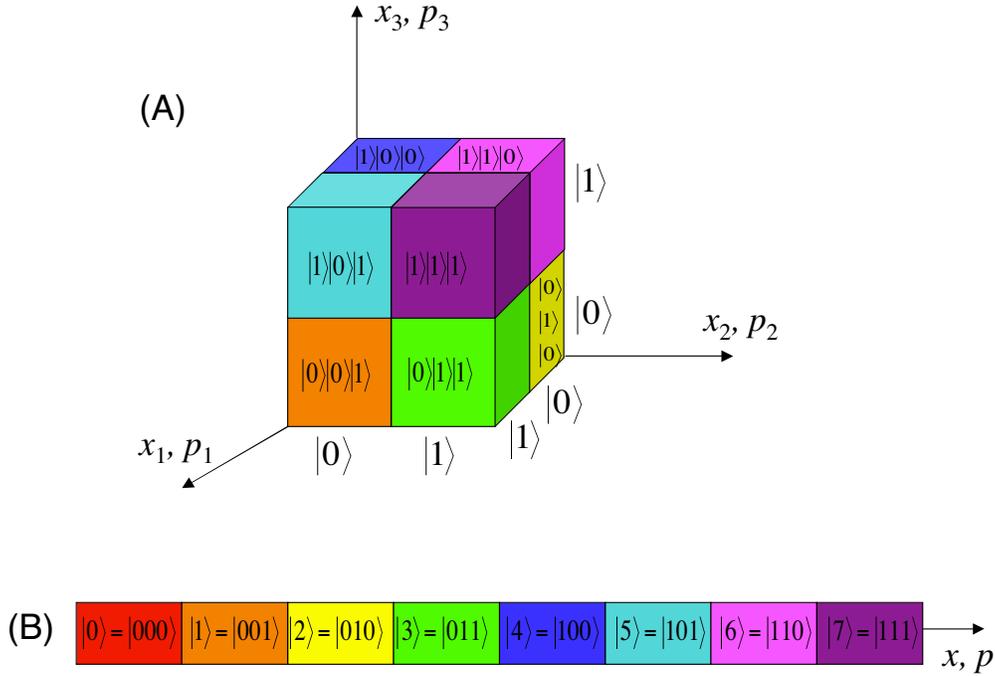}
\end{center}
\vspace{18pt}
\caption{
{\em Using many degrees of freedom to save resources.} Orthogonal
basis states for an eight-dimensional Hilbert space depicted
schematically as nonoverlapping phase-space cells in the phase space
of three degrees of freedom (A), each of which uses an action $\sim
2h$, or in the phase space of a single degree of freedom (B). Phase
space is pictured at half its actual dimension by letting the axes
represent both the position and momentum co\"ordinates for a degree
of freedom; one can think of the axes as measuring the amount of
action used by a degree of freedom. To accommodate the eight states,
the single degree of freedom requires three times as much action as
does each of the three degrees of freedom.  If one adds degrees of
freedom to (A), the phase-space volume---and hence the Hilbert-space
dimension---doubles as each degree of freedom is added and thus grows
exponentially with the number of degrees of freedom, whereas the
physical resources grow linearly with the number of degrees of
freedom and thus logarithmically with the Hilbert-space dimension.
The result is a scalable resource requirement.  In contrast, for the
single degree of freedom in (B), the required resources grow linearly
with phase-space volume and Hilbert-space dimension; to achieve the
same Hilbert-space dimension as for the scalable case requires
physical resources that are exponentially larger.  As shown, the
basis states for both situations can be labeled either by unary or
binary numbers, this being an example of the fungibility of Hilbert
spaces.  The labeling, however, cannot alter the physics: the single
degree of freedom is a {\em physically\/} unary realization of the
Hilbert space, which uses exponential resources asymptotically,
whereas the multiple degrees of freedom in (A) provide a {\em
physically\/} binary realization of the same Hilbert space, which
uses resources efficiently.  The compact phase space achieved in (A)
also aids in suppressing decoherence, as discussed in
Sec.~\ref{sec:stability}.}
\label{fig1}
\end{figure}

It is useful to summarize this simple result, as it is the foundation
for all our further conclusions.  The physical resources are the
quantities that label the axes of a (generalized) phase space that
has two axes for each degree of freedom.  The number of Hilbert-space
dimensions available for a computation is proportional to the total
phase-space volume.  If the number of degrees of freedom grows
linearly in $N$, the phase-space volume needed to accommodate the
Hilbert-space dimension can be fitted into a hypercube in phase space
without requiring an exponentially increasing contribution along any
direction in phase space.  In contrast, if the number of degrees of
freedom grows more slowly than linearly in $N$ (within the
logarithmic corrections), some phase-space direction must supply an
exponentially increasing amount of the corresponding physical
resource.  This simple argument is depicted schematically in
Fig.~\ref{fig1}.

To formulate a more precise statement, we specialize to the case of
$T$ identical degrees of freedom, each of which supplies an action
$A$. In this situation, the total number of Hilbert-space dimensions
satisfies $(A/h)^T\sim2^N$, which gives
\begin{equation}
A/h\sim2^{N/T}\;.
\end{equation}
In order to avoid an exponential resource demand, $A/h$ must grow
polynomially with $N$ [23], which means that the number of degrees of
freedom increases as [24]
\begin{equation}
T\sim{N\over\log{\bf P}(N)}\;,
\label{eq:quasilinear}
\end{equation}
where ${\bf P}(N)$ is a function bounded above by a polynomial.  We
say that $T$ grows {\em quasilinearly\/} with $N$ and that the system
is {\em scalable}, having a {\em scalable tensor-product structure}.

For comparison with our analysis of quantum fields, it is instructive
to distinguish three cases:
\begin{enumerate}
\begin{item}
$T$ grows more slowly than linearly with $N$.  If $T$ grows
quasilinearly, as in Eq.~(\ref{eq:quasilinear}), then $A/h\sim{\bf
P}(N)$, and the system is scalable.  If $T$ grows more slowly than
quasilinearly with $N$, $A/h$ grows exponentially with $N$, leading
to an exponential demand for physical resources.
\end{item}
\begin{item}
$T$ grows faster than linearly with $N$.  Since $A/h$ goes to one as
$N$ increases, the present analysis in terms of independent degrees
of freedom breaks down and should be replaced by a counting of the
excitations of a quantum field, which we give in Sec.~\ref{sec:Fock}.
\end{item}
\begin{item}
$T=N/\log D$ grows strictly linearly with $N$.  For $D<2$, the
present analysis breaks down, and we again need the analysis of
quantum fields to reach a sensible conclusion.  For $D\ge2$, each
degree of freedom is a $D$-level system, i.e., a {\em qudit\/}
instead of a qubit.  Though this is a special case of quasilinear
growth in which ${\bf P}(N)=D$, we separate it off for separate
analysis.  It is the most important scalable case because the action
supplied by each system, $A/h\sim D$, is independent of problem size.
Scaling is achieved simply by adding degrees of freedom, without
having to change the Hilbert-space dimension supplied by each degree
of freedom.  We say that this kind of system is {\em strictly
scalable\/} and has a {\em strictly scalable tensor-product
structure}.  Most quantum computing proposals are of this sort.
\end{item}
\end{enumerate}

Had we focused on the total action resource,
\begin{equation}
TA/h\sim T2^{N/T}\;,
\label{eq:totalaction}
\end{equation}
instead of on the action resource per degree of freedom, we would
have reached the same conclusions regarding scaling.  The total
action resource is more akin to the resource quantities that arise in
our analysis of quantum fields.  For a scalable system, it grows as
$TA/h\sim N{\bf P}(N)/\log{\bf P}(N)$; only for strictly scalable
systems is the total action resource linear in $N$.

\subsection{Quantum Computing in a Single Atom}
\label{sec:singleatom}

An illuminating extreme example of the nonscalable systems in case~1
is the attempt to implement quantum computing in a single atom
[17,\hspace{2.5pt}25,\hspace{2.5pt}26], fixed molecule [18], or large spin [19].
Advances in laser spectroscopy with ultrashort pulses have allowed
researchers to manipulate and measure the electronic wave function in
an atom [27] or both electronic and rotational/vibrational wave
functions in a molecule [28] with exquisite precision. It is natural
to wonder whether these tools for coherent control of quantum states
can be applied to quantum computing.

For illustration, consider the simplest hypothetical model, quantum
computing in a hydrogen atom.  Characteristic atomic units of length,
momentum, and energy are formed from the physically important
constants: the electron charge and mass, $e$ and $m$, and the quantum
of action, $\hbar$.  If we ignore spin, Bohr's formula for quantizing
the action gives the familiar expressions for the energy, radius, and
momentum of a stationary state with principle quantum number $n$,
\begin{equation}
E_n=-{1\over {2n^2}}\,{{e^2} \over {a_0}}\;,\quad
r_n=n^2a_0\;,\quad
p_n={1 \over n}\,{\hbar  \over {a_0}}\;,
\end{equation}
where $a_0 = \hbar^2 / m e^2$ is the Bohr radius.  The dimension of
the Hilbert space spanned by all bound states from the ground state
up to a maximum principle quantum number $n$ is
\begin{equation}
\sum_{k=1}^n \sum_{l=0}^{k-1}
(2l+1)\sim {1\over 3}n^3 \sim \left( {r_np_n \over \hbar } \right)^3\;.
\end{equation}
The final expression is of just the form we expect.  Without spin the
internal states of the hydrogen atom have three degrees of freedom,
signaled by the 3 in the exponent and corresponding to the three
co\"ordinates of relative motion of the electron and proton. Each
degree of freedom is allotted an action $A\sim r_n p_n$, which provides
enough phase space for $\sim A/h$ orthogonal states in Hilbert space.

Demanding that the atomic Hilbert space have a dimension $2^N$
requires that the radial co\"or\-di\-nate scale as $r_n\sim
2^{2N/3}a_0$. The exponential growth of this co\"or\-di\-nate with
problem size implies that quantum control in a single atom {\em
cannot\/} be used for scalable quantum computation.  For instance, to
implement a quantum computation requiring $N=100$ qubits, the atomic
radius must be $r_n\sim 10^{20}a_0=6\times 10^6\,$km, about 5 times
the diameter of the Sun.

A single atom is an example of a ``physically unary'' quantum
computer, having a limited natural tensor-product structure provided
by the small number of physical degrees of freedom.  Similar poor
scaling will be seen in any implementation consisting of a single
particle, a single atom, or a single molecule consisting of a fixed
number of atoms.  The fungibility of Hilbert spaces means that one
can impose an artificial tensor-product structure on the Hilbert
space of these systems, equivalent to that of qubits, but this does
not obviate the need to provide the physical resources to generate
orthogonal quantum states. Without a scalable tensor-product
structure corresponding to a division into physical degrees of
freedom, one or more of the physical co\"ordinate axes must blow up
exponentially with problem size, meaning that these systems are not
suitable for scalable quantum computation.

This should be contrasted with quantum computing using multiple
atoms, containing a physical tensor product structure, such as in an
ion trap [29].  Quantum information is stored in two sublevels  of
each of the ion's ground states and manipulated with a limited number
of vibrational states.  A Hilbert space of 100 qubits requires 100
ions in their ground states occupying 100 local positions.  Neither
the internal nor the external degrees of freedom of the atoms require
physical resources that grow exponentially in order to accommodate a
$2^N$-dimensional Hilbert space.

We now need to extend the lessons of this section to the more general
case of quantum fields.  In that context, the notion of degrees of
freedom is generally not well defined, though in some circumstances
it re\"emerges as a useful concept.  This more general analysis
allows us analyze the cases that we were unable to treat properly
above.  Readers not interested in these details can skip the next
section with little loss of continuity.

\section{FOCK-SPACE ANALYSIS OF RESOURCE \\ REQUIREMENTS}
\label{sec:Fock}

\subsection{Resources in Fock Space}
\label{sec:Fockresources}

We now consider a quantum field to be the basic physical system. The
state of a single particle, i.e., a single quantum of excitation of
the field, is described in a Hilbert space that is a tensor product
of a $K$-dimensional Hilbert space for the particle's external
degrees of freedom (e.g., translational motion in three spatial
dimensions) and a $D$-dimensional Hilbert space for the internal
degrees of freedom (e.g., spin).  The single-particle states
represent different configurations of the quantum field, analogous to
wave functions, and are often called field ``modes.''  The total
number of modes is $M=KD$.  Given the single particle space---``first
quantization''---we can define the many-body system through the
Fock-space construction---``second quantization.''  Fock space is
spanned by orthonormal {\em Fock states}, which are specified by
giving the number of particles in each of the single-particle states.

The physical resources are the total number of particles, $L$, and
the numbers of external and internal single-particle states, $K$ and
$D$.  We consider three kinds of systems: bose and fermi systems, and
systems where each external state contains at most one particle.  In
the last of these, the particles are distinguished by the label for
the external state and thus act like ``distinguishable'' particles.
When $L=K$, each ``distinguishable'' particle has available $D$
internal states; hence this case reduces to $T=L=K$ particle degrees
of freedom, each with $A/h=D$ levels, i.e., a quantum computer
consisting of $L$ qudits.

For quantum fields, field and particle degrees of freedom are
slippery concepts, which become rigorous only in special cases, such
as the case of ``distinguishable'' particles just mentioned.  In
classical physics, particles and fields are both described by pairs
of canonical coordinates, with the number of pairs determining the
number of degrees of freedom.  Thus a point particle moving in three
dimensions has access to three degrees of freedom, and a vibrating
string of limited bandwidth has access to a set of fundamental modes,
each of which is a degree of freedom.  The complementary particle and
field aspects of a quantum field mean that physical degrees of
freedom cannot generally be defined rigorously for quantum fields,
since a rigorous definition requires that the overall Hilbert space
be a tensor product of the Hilbert spaces for the individual degrees
of freedom.  The particle degrees of freedom of a quantum field come
from a particle's ability to occupy various single-particle states,
but the restrictions set by particle indistinguishability mean that
Fock space is not a tensor product of particle Hilbert spaces.  The
field degrees of freedom arise from the different numbers of
particles that can occupy a single-particle state, or field mode.
Although the entirety of Fock space is a tensor product of the
field-mode Hilbert spaces, each spanned by particle-number states,
the subspaces we are considering, which have no more than a fixed
number of particles, are not.  For example, for a bose field
containing exactly $L$ particles, the states where any given mode
contains all of the particles is in the subspace, but the tensor
product of these states, where all modes contain $L$ particles,
clearly is not.

The particle and field degrees of freedom of a quantum field can,
nonetheless, be serviceable approximate concepts.  It is useful to
think in terms of particle degrees of freedom when the number of
modes per particle, $M/L$, is large; we can then think of the $L$
particles as effective degrees of freedom.  Likewise, field degrees
of freedom are a useful approximate concept when the number of
particles per mode, $L/M$, is large; in this case we can think of the
$M$ modes as effective degrees of freedom.  Outside these asymptotic
regimes, particle and field aspects are both important, and the
degrees of freedom are less useful concepts.  Of course, for
fermions, field degrees of freedom are never a useful concept,
because the possible field excitations are so restricted by the Pauli
exclusion principle.

For bosons, the physical resources can be interpreted in terms of a
phase-space picture.  The electromagnetic field provides a familiar
example: the field modes give the possible states for a photon, and
the population of the modes by photons describes the amplitudes of
the electric and magnetic fields.  The total number of modes, $M$, is
proportional to the phase-space volume used by a single particle; it
characterizes how ordinary space and particle momentum (wave number)
and also internal states like photon polarization are used as
resources.  The number of particles, $L$, is proportional to the
volume used in the phase space of the bose field; it characterizes
how field strength is used as a resource.  Because of the exclusion
principle, only the particle aspect of this phase-space picture works
for fermions, but that is sufficient for our considerations; since
$L\le M$, the number of modes is always the important resource for
fermions.

Quantum entanglement is only defined for Hilbert spaces that have a
rigorous tensor-product structure in terms of subsystems.  Thus the
structure of Fock space as a tensor product of field-mode Hilbert
spaces has important implications for entanglement: entanglement
among field modes is always well defined, but particle entanglement,
along with particle degrees of freedom, can be defined rigorously
only in special cases [19], an example being distinguishable
particles with $L=K$.  Note that when the field modes share just a
single particle, mode entanglement is nothing more than the
second-quantized version of a simple superposition state in the
language of first quantization.  These superposition states, e.g.,
the state of a single photon after it passes though a beam splitter,
are indeed entangled states and can be used as an entanglement
resource in protocols such as teleportation [30].

The Fock-space construction in hand, we proceed to counting
Hilbert-space dimensions and analyzing resource requirements for the
three kinds of systems.  The counting is equivalent to calculating
the entropy of a microcanonical ensemble in which all particles carry
the same energy.

\subsection{Scaling in Bose Systems}
\label{sec:bose}

The dimension of the Hilbert space for $L$ bosons occupying $M$ modes
is
\begin{equation}
\Omega_B={(M+L-1)!\over(M-1)!\,L!}\;.
\end{equation}
This expression is invariant under the exchange $L\leftrightarrow
M-1$; i.e., we can effectively exchange the roles of particles and
modes in counting the number of orthogonal Fock states.  For bosons
it is often useful to consider the situation where the number of
particles, instead of being fixed, can vary from zero to a maximum
number $L_{\rm max}$.  The corresponding Hilbert-space dimension can
be obtained from $\Omega_B$ by increasing the mode number by 1, i.e.,
by imagining that there is an additional ``phantom'' mode that soaks
up the extra particles:
\begin{equation}
\Omega'_B={(M+L_{\rm max})!
\over M!\,L_{\rm max}!}\;.
\end{equation}
The particle-mode symmetry in this case is even simpler:
$L_{\rm max}\leftrightarrow M$.

We consider whether this many-body system can support scalable
quantum computation by examining the asymptotic behavior of
$\Omega_B$ (or $\Omega'_B$) in various cases.

\begin{enumerate}
\begin{item}
$L$ fixed, $M$ grows: $2^N=\Omega_B\sim M^L/L!\,$.  Particle degrees
of freedom predominate.  The system is not scalable because the
number of modes must grow exponentially with $N$.
\end{item}

\begin{item}
$M$ fixed, $L_{\rm max}$ grows: $2^N=\Omega'_B\sim L_{\rm
max}^M/M!\,$.  Field degrees of freedom predominate.  The system is
not scalable because the number of particles must grow exponentially
with $N$.
\end{item}

\begin{item}
Both $L$ and $M$ grow:
\begin{equation}
2^N=\Omega_B\sim
\left(1+{L\over M}\right)^M\left(1+{M\over L}\right)^L\;.
\end{equation}
The first term represents field degrees of freedom, and the second
term represents particle degrees of freedom.  In this asymptotic
regime the particle-mode symmetry reduces to $L\leftrightarrow M$.
Again there are three cases.

(i)~$M$ grows faster than linearly with $L$: $N=\log\Omega_B\sim
L\log(M/L)\Longrightarrow M\sim L2^{N/L}$ [this has the same form as
the total action resource in Eq.~(\ref{eq:totalaction})].  Particle
degrees of freedom predominate, with $L$ being an effective number of
degrees of freedom and $M$ being the resource that must be
constrained.  To be consistent with this case, $L$ must grow more
slowly than linearly with $N$.  As in our degrees-of-freedom
analysis, if $L\sim N/\log{\bf P}(N)$ grows quasilinearly with $N$,
the growth of $M\sim N{\bf P}(N)/\log{\bf P}(N)$ leads to a scalable
resource requirement.  If $L$ grows more slowly than quasilinearly
with $N$, then $M$ grows exponentially with $N$, giving a nonscalable
resource requirement.

(ii)~$L$ grows faster than linearly with $M$: $N=\log\Omega_B\sim
M\log(L/M)\Longrightarrow L\sim M2^{N/M}$.  Field degrees of freedom
predominate, with $M$ being an effective number of degrees of freedom
and $L$ being the resource that must be constrained.  We reach the
same conclusions as for (i), but with $L$ and $M$ reversed.

(iii)~$L=\mu M$, $\mu$ (constant) being the average number of
particles per mode:
\begin{equation}
2^N=\Omega_B\sim(1+\mu)^M(1+\mu^{-1})^L=2^{MS(\mu)}=2^{LS(1/\mu)}\;.
\end{equation}
Here $S(\mu)\equiv-\mu\log\mu+(1+\mu)\log(1+\mu)$ is the entropy (in
bits) of a field mode containing on average $\mu$ quanta.  The
available Hilbert-space dimension is that of $M$ degrees of freedom,
each with $2^{S(\mu)}$ levels, or $L$ degrees of freedom, each with
$2^{S(1/\mu)}$ levels, in accordance with the particle-mode symmetry.
This case is {\em strictly\/} scalable, as both $M$ and $L$ grow
linearly with $N$. For $\mu\gg1$, where field degrees of freedom
predominate, the counting, $\Omega_B\sim\mu^M$, reduces to that of
$M$ modes, each with $\mu$ levels.  For $\mu\ll1$, particle degrees
of freedom predominate, and the counting, $\Omega_B = (\mu^{-1})^L$,
reduces to that of $L$ particles, each with $\mu^{-1}$ levels; in
this asymptotic regime, we recover the simple degrees-of-freedom
analysis for the bose particles, each of which has access to a
phase-space volume proportional to $\mu^{-1}$.
\end{item}

\end{enumerate}

Examples of these different scenarios have been explored in the
literature.  Physically unary systems are a special instance of
case~1 with a single particle ($L=1$) or of case~2 with a single mode
($M=1$).  In a single-particle Fock space, we have $2^N=\Omega_B=M$,
and there are two interesting possibilities: $D=1$, $K=M$ corresponds
to single-photon optics [31,\hspace{2.5pt}32], whereas $K=1$, $D=M$ corresponds to
an $M$-level system like an atom [17].  Both of these require an
exponential number of modes and the associated physical resources.
The case of many bodies occupying a single mode (case~2 with $M=1$)
corresponds to quantum optics in a single-mode cavity; though this
system has a large number of ``nonclassical'' states (e.g., squeezed
states), the particle number must scale exponentially,
$2^N=\Omega'_B=L_{\rm max}+1$.

Closely related to unary systems with a single particle are
implementations of quantum algorithms that use superposition and
interference of classical linear waves.   Classical linear optics
(electromagnetic waves) provides an example that can be easily
implemented in the laboratory.  The wave amplitudes are described in
a complex vector space, just like the Hilbert space of a quantum
system, so it might appear that such classical-wave processors are
candidate quantum computers.  The problem is that they will {\em
always\/} scale poorly when the necessary physical resources are
taken into account.  A classical wave is essentially a many-particle
copy of a single-particle wave function.  The linear-optics
transformations of a classical wave are in one-to-one correspondence
with the unitary transformations of the single particle wave
function.  The single photon has only three motional degrees of
freedom and one polarization degree of freedom.  Thus a
classical-wave computation requires an exponential number of modes in
the single-particle phase space [33], a demand inherited from a
single-particle unary machine [34].  In addition, since the
transformations required for a computation generally populate an
exponential number of distinguishable modes, a classical-wave
computation requires an additional exponential overhead in particle
number (field strength) if all the populated modes are truly
classical throughout the computation.  This additional overhead can
be avoided if one drops the demand for classical waves at all
intermediate stages of the computation.

A compelling illustration of the physical-resource demands in
classical linear-optical implementations was provided by Bhattacharya
{\em et al.}\ [35] in a simulation of Grover's algorithm for
searching a database.  The database entries were represented by the
dif\-frac\-tion-limited transverse modes of a laser beam.
Classical-wave interference leads to effective amplification of the
sought-after mode, as Grover's algorithm predicts.  As the database
grows with the corresponding number of qubits, however, the waist
diameter of the beam must grow exponentially and would reach the size
of the visible universe for $\sim 220$ qubits [36].  These same
resource demands are seen in single-photon (unary) linear
interferometers used to simulate quantum algorithms [31,\hspace{2.5pt}32].

The classical-wave example demonstrates that just having the
necessary scalable Hilbert space is not sufficient to ensure scalable
quantum computing.  Classical waves are coherent states with a large
mean particle number; the restriction to linear-optical
transformations means that the field always stays within the
coherent-state sector, never exploring the multitude of
``nonclassical'' many-body states. Whereas the Hilbert space of this
many-boson, many-mode system can be made sufficiently large without
exponential use of physical resources [case~3(iii)], the classical
waves explore only a tiny portion of the available states. In doing
so, classical-wave vector spaces end up demanding exponential
resources to keep up with the quantum Hilbert space.

In contrast to classical waves, examples of bose systems that can
take advantage of the favorable scaling of case~3(iii)---in
particular, a proposal to use nonlinear optics as a source of
interactions between pairs of photons [37]---were among the earliest
proposals for quantum computation.  More recently and more
surprisingly, Knill, Laflamme, and Milburn [38] have demonstrated
that just with {\em linear\/} optical unitary transformations---i.e.,
one-body transformations and no interactions between photons---one
can implement scalable quantum computing, provided one has access to
nonclassical field inputs and measurements of photon number, both of
which take the field out of the coherent-state sector.  In contrast,
it has been shown [39,\hspace{2.5pt}40] that if one starts in a state with Gaussian
statistics and has access only to manipulations within the so-called
``Clifford semigroup'' [40], which includes linear optics, squeezing,
fast feedforward, and generalized measurements of canonical
observables, but does not include photon counting, the result can be
classically simulated and thus does not correspond to universal
quantum computation.   These examples demonstrate the subtlety of
determining whether a given system has access to {\em arbitrary\/}
states in Hilbert space.

\subsection{Scaling in Fermi Systems}
\label{sec:fermi}

We now consider $L$ fermions distributed among $M$ modes ($L \le M$).
The number of distinguishable configurations gives rise to a Hilbert
space of dimension
\begin{equation}
\Omega_F={M!\over L!\,(M-L)!}\;.
\end{equation}
Fermi systems exhibit a particle-hole symmetry, $L\rightarrow M-L$.
As with the bose case, we look at the asymptotic behavior to learn how
the resource requirements scale.

\begin{enumerate}

\begin{item}
$L$ fixed, $M$ grows: $2^N=\Omega_F\sim M^L/L!\,$.  This is
equivalent to case~1 for bosons, because the particles occupy the
modes sparsely.  We reach the same conclusion, i.e., this case is not
scalable.
\end{item}

\begin{item}
Both $L$ and $M$ grow:
\begin{equation}
2^N=\Omega_F\sim
\left({1\over1-L/M}\right)^{M-L}\left({M\over L}\right)^L\;,
\quad
L\le M\;.
\end{equation}
Now there are two subcases.

(i)~$M$ grows faster than linearly with $L$: $N=\log\Omega_F\sim
L\log(M/L)$.  This is identical to case~3(i) for bosons, since the
particles are sparse, and we reach the same conclusions, i.e.,
scalability if $L$ grows quasilinearly with $N$, but not otherwise.

(ii)~$L=\mu M$, $\mu\le1$ (constant) being the average number of
particles per mode and $1-\mu$ the average number of holes per mode:
\begin{equation}
2^N=\Omega_F\sim
(1-\mu)^{-(1-\mu)M}\mu^{-\mu M}=2^{MH(\mu)}\;.
\end{equation}
Here $H(\mu)\equiv-\mu\log\mu-(1-\mu)\log(1-\mu)\le1$ is the binary
Shannon entropy corresponding to fraction $\mu$.  The dimension of
this Hilbert space is like that of $M$ degrees of freedom, each with
$2^{H(\mu)}$ levels.  The particle-hole symmetry becomes $\mu
\leftrightarrow 1-\mu$.  This system is {\em strictly\/} scalable
since both $M$ and $L$ grow linearly with $N$.  We recover an
effective picture of particle degrees of freedom for $\mu\ll1$, where
$\Omega_F\sim(\mu^{-1})^L$, and of hole degrees of freedom for
$1-\mu\ll1$, where $\Omega_F\sim [(1-\mu)^{-1}]^{M-L}$. The largest
Hilbert space arises for equal numbers of particles and holes,
$\mu=1/2$, where $\Omega_F\sim2^M=4^L$.
\end{item}

\end{enumerate}

Examples of scalable fermi systems [case~2(ii)] have been
investigated.  Bravyi and Kitaev [41] showed that there is a
universal gate set that consists of linear transformations together
with a transformation coming from an interaction that is quartic in
field amplitudes.  In contrast to the bose case, the noninteracting
fermi gas with measurements that count particles does {\em not\/}
allow for universal quantum computation [42,\hspace{2.5pt}43].  Once again we see
that access to a scalable Hilbert space is necessary, but not
sufficient for performing quantum computation.

\subsection{Scaling for ``Distinguishable'' Particles}
\label{sec:distinguishable}

When there is no more than one particle in each external state, the
particles are effectively distinguishable. We assume $D\ge2$ since
$D=1$ reduces to the fermi case.  The number of configurations is
\begin{equation}
\Omega_D={K!\over L!\,(K-L)!}D^L\;,\quad L\le K.
\end{equation}
As promised, we recover the qudit case when $L=K$, but we now have
the freedom to explore the intermediate possibilities that arise when
there are not enough particles to fill each of the external states,
i.e., $1\le L<K$.  Here $K$ plays the role of the number of degrees
of freedom in our simple degrees-of-freedom analysis, and $D$ plays
the role of $A/h$.  In contrast to the first-quantized picture, here
$A/h=D$ gets raised to the power $L$, not $K$, when $L<K$, because
not all the external states are occupied.  This allows us to deal
with the cases that we were unable to handle previously because we
are now properly taking into account the resources required by
unoccupied modes, i.e., vacuum.

We consider the number of internal states to be fixed in our analysis
of the asymptotics, because the case where $D$ grows has already been
dealt with in our simple degrees-of-freedom analysis.  With this
assumption, the asymptotic analysis goes as follows.

\begin{enumerate}
\begin{item}
$L$ fixed, $K$ grows: $2^N=\Omega_D\sim(KD)^L/L!\,$. This is
equivalent to case~1 for bosons and fermions, because the particles
sparsely occupy the modes.  The system is not scalable because the
number of single-particle states must grow exponentially with $N$.
\end{item}

\begin{item}
Both $L$ and $K$ grow:
\begin{equation}
2^N=\Omega_D\sim
\left({1\over1-L/K}\right)^{K-L}\left({KD\over L}\right)^L\;,
\quad
L\le K\;.
\end{equation}
Now there are two subcases.

(i)~$K$ grows faster than linearly with $L$: $N=\log\Omega_D\sim
L\log(KD/L)$.  This is a realization of case~2 in our simple
degrees-of-freedom analysis, which we were unable to analyze because
in that treatment we could not account for the resources required by
unoccupied modes.  Since the particles sparsely occupy the modes,
this case is identical to case~3(i) for bosons and case~2(i) for
fermions, i.e., scalable if $L$ grows quasilinearly with $N$, but not
otherwise.

(ii)~$L=\mu K$, with $\mu\le1$ (constant):
\begin{equation}
2^N=\Omega_D\sim
(1-\mu)^{-(1-\mu)K}\mu^{-\mu K}D^L=2^{K[H(\mu)+\mu\log D]}\;.
\label{eq:d2ii}
\end{equation}
The system is {\em strictly\/} scalable with $K$ and $L$ linear in
$N$: $L= \mu K=N/[H(\mu)/\mu+\log D]$.  This provides the correct
treatment of the remaining unanswered question in case~3 of the
degrees-of-freedom analysis.
\end{item}

\end{enumerate}

Though all these specific cases are tedious to analyze, there is a
payoff, for they come together in a fundamental requirement for a
many-body system to be a scalable quantum computer: {\em scalability
requires that the number of particles or the number of modes,
whichever (or both) acts as the effective number of degrees of
freedom, must grow quasilinearly with the equivalent number of
qubits, $N$; if the effective number of degrees of freedom grows more
slowly than quasilinearly in $N$, the complementary resource set
demands an exponential supply of physical resources.}  This
requirement is the analogue of our conclusion that a set of degrees
of freedom must have a scalable tensor-product structure.  The
many-body analogue of strict scalability is that both $L$ and $M$
grow strictly linearly with $N$, this being the only case where all
resources grow linearly with~$N$.

\section{OTHER REQUIREMENTS FOR A SCALABLE \\ QUANTUM COMPUTER}
\label{sec:otherreq}

\subsection{DiVincenzo Requirements}
\label{sec:DiVincenzo}

So far we have analyzed one necessary condition for a scalable
quantum computer, based on the need to avoid an exponential demand
for physical resources.  We have been careful to emphasize that this
requirement is necessary, but by no means sufficient.  To see how the
physical-resource requirement is related to other requirements for
implementing a universal quantum computer, it is instructive to
consider the list of five requirements laid down by DiVincenzo [3],
which we have modified slightly for our purposes.

\begin{enumerate}
\begin{item}
{\em Scalability.} A scalable physical system with well
characterized parts, usually qubits.
\end{item}
\begin{item}
{\em Initialization.} The ability to initialize the system
in a simple fiducial state.
\end{item}
\begin{item}
{\em Control.} The ability to control the state of the
computer using sequences of elementary unitary operations chosen
from a set of universal gates.
\end{item}
\begin{item}
{\em Stability.} Long relevant decoherence times, much longer
than the gate times, together with the ability to suppress
decoherence through error correction and fault-tolerant computation.
\end{item}
\begin{item}
{\em Measurement.} The ability to read out the state of the computer
in a convenient product basis called the computational basis.
\end{item}
\end{enumerate}

The first item in DiVincenzo's list posits that a scalable quantum
computer must be made up of parts with a strictly scalable
tensor-product structure.  Where does this requirement come from?  Is
it a prior requirement, independent of the other items in the list,
or is it needed for initialization, control, stability, and efficient
measurement?  We argue here that a strictly scalable tensor-product
structure is a {\em prior\/} requirement, above all others: in
providing the primary resource of Hilbert-space dimension, a scalable
system is necessary to avoid an exponential demand for physical
resources, and a strictly scalable system is needed to constrain the
demand for resources to grow as slowly as possible, i.e., linearly in
the equivalent number of qubits.

DiVincenzo's further requirements come into play once one has dealt
with the resource issue.  We suggest that a strictly scalable
tensor-product structure makes it easier to achieve the control and
stability requirements---so much easier, in fact, that one can regard
a strictly scalable tensor-product structure as essential in practice
for these two requirements.  We turn now to a discussion of how the
control and stability requirements are related to DiVincenzo's first
requirement, also touching on the question of quantum information
processing using mixed states and the thorny question of the role of
entanglement in quantum computing.  We do not consider measurements
issues explicitly, except as they arise in our discussion of the need
for many measurements to read out the output of a mixed-state
computer.

\subsection{Control}
\label{sec:control}

Control of a quantum computation is accomplished via some set of
elementary ``universal" operations.  In the quantum circuit model,
these can be a finite set of one- and two-qubit quantum logic gates
(unitary operators) [44] or an equivalent set of Hamiltonians that
generate the one- and two-qubit dynamics [45,\hspace{2.5pt}46].  Alternatively,
quantum algorithms can be implemented through a series of projective
quantum measurements and classical control [4--6].  These schemes
assume a tensor-product structure, usually a qubit decomposition. The
qubit structure makes physical implementation of the elementary
operations straightforward in principle; the coupling to the system
needs to isolate either a single qubit or a pair of interacting
qubits.  Though many experimentalists will bridle at our use of
``straightforward,'' the control issues in systems without a
tensor-product structure are far more serious, as noted by Ekert and
Jozsa [47].

Consider, for example, quantum control of a unary system such as a
single atom [17] or a large spin [19].  One control strategy is to
map the one- and two-qubit gates onto the $2^N$ levels of the unary
system.  Even the simplest of the required gates, however, is
difficult to implement in terms of operators that are physically
relevant to the unary system.  For instance, in a three-qubit system,
the gate $\sigma_x^{(1)}\otimes I^{(2)} \otimes I^{(3)}$ generates a
bit flip on the first qubit.  Written in an 8-dimensional unary
representation, this gate involves transitions between the $m$th
level to the $(m \pm 4)$-th level, and all four transitions have the
same strength.  This involves coupling to the entire unary system, in
contrast to the single-qubit coupling that is natural in a system
made of qubits.  The same problem arises for any mapping onto a
``virtual subsystem'' [16].  In practice, control of physically unary
systems would be achieved by coupling directly to each level and by
pairwise transitions between levels; since this requires access to a
$\sim2^{2N}$ control parameters, it is not scalable.

The relative ease with which a quantum system built out of subsystems
can be controlled can be understood in terms of degrees of freedom.
The physical quantities that quantify the resources for a degree of
freedom provide the connection to the external world; precisely
because they are physical observables, these physical quantities are
available for building Hamiltonians that are controlled by an
external classical apparatus.  This allows the experimentalist to
manipulate an exponential number of probability amplitudes with a
polynomial number of gate operations.

\subsection{Stability}
\label{sec:stability}

A scalable tensor-product structure aids in suppressing decoherence
and is probably essential for implementing quantum error correction
and fault-tolerant quantum computation.  The simplest analysis of the
decoherence of quantum states that are widely separated in phase
space gives a decoherence rate that is proportional to the square of
the phase-space distance between the states [48,\hspace{2.5pt}49].  Our phase-space
picture of the physical resources used by a quantum computation (see
Fig.~\ref{fig1}) shows that a qubit-based scalable quantum computer
occupies a region of phase space that looks roughly like a
$2N$-dimensional hypercube with side lengths independent of the
number of qubits; the greatest distance between any two states in the
accessible region of phase space is thus proportional to $\sqrt{2N}$.
In contrast, in a unary system, where one degree of freedom bears the
entire burden of the exponential increase of Hilbert-space dimension
with problem size, the greatest distance between states grows at
least as fast as $2^{N/2}$, i.e., exponentially with the equivalent
number of qubits. This sharp difference suggests that a scalable
tensor-product structure can play an important role in reducing
decoherence.  We emphasize that this argument is based on a very
crude model of decoherence.  Decoherence is not only highly
system-specific, but difficult to characterize simply even for
specific systems [50].  The significance of the argument is to
suggest that a system whose accessible states are compactly arranged
in phase space will not decohere faster than one whose states are
distant in phase space and, under appropriate circumstances, will
decohere much more slowly.

Once decoherence and noise in a physical system have been reduced
below the error threshold for fault-tolerant quantum computation
[10--13], quantum error correction [7--9] can be used to suppress
errors sufficiently to perform arbitrarily long computations.
Error-correction protocols cannot correct all errors. Instead they
seek to correct the most probable errors, where what is most probable
depends on the error mechanisms appropriate for a specific physical
system; examples of such dominant errors include errors that act
independently on individual qubits (though we refer to qubits, qudits
could be used just as well) or errors that are correlated over many
qubits.  The most probable errors define an ``error algebra'' [9] of
errors to be corrected.  To detect and correct these errors, one
encodes ``logical qubits'' into carefully chosen two-dimensional
subspaces of several qubits.  A good code is one such that the
generators of the error algebra map the code subspace unitarily into
mutually orthogonal subspaces.  One is thus able to diagnose the
error and correct it without destroying the encoded quantum
information.  We argue that these error-correction protocols require
a scalable tensor product structure.

One set of errors that must be corrected consists of the inevitable
imperfections in the quantum logic gates.  Control of a set of qubits
(or qudits) can be accomplished by a set of quantum logic gates whose
number is polynomial in $N$.  In contrast, as noted above, the
natural couplings to a system that has no physical tensor-product
structure are direct couplings to individual levels and pairwise
transitions between levels.  Arbitrary unitary operations can be built
out of these elementary interactions, but since there are $\sim2^{2N}$
elementary interactions, errors in them will lead to an exponentially
large error algebra that contains essentially all errors, thus making
error correction impossible.

Even if we had a more efficient scheme for constructing arbitrary
unitaries, the known error-correction schemes still require a
tensor-product structure.  Error-correcting codes work by channeling
the entropy introduced by noise and decoherence into ancillary
subsystems (typically a number of qubits), which are reinitialized in
a pure state for subsequent rounds of error correction.  It is
difficult to see how this could be managed in a system not made of
parts that can be used as the ancillary subsystems.  In particular,
it is difficult to see how a virtual subsystem within a Hilbert space
without a tensor-product structure could be reinitialized---or even
how fresh virtual subsystems could be introduced.

\subsection{Mixed-State Quantum Computing}
\label{sec:mixedstates}

Our analysis of physical-resource requirements assumed implicitly
that the quantum system is described by a pure quantum state.  Yet
the system that has implemented the most sophisticated
quantum-information-processing protocols is liquid-state nuclear
magnetic resonance (NMR) [51--53], where the nuclear spins that act
as qubits are described by a highly mixed state. A mixed-state
quantum information processor can have a strictly scalable
tensor-product structure, as do the nuclear-spin qubits of NMR, yet
still require exponential resources because of the mixed nature of
the quantum state.  The problem is one of initialization. When the
physical system is initialized in a mixed state, it has some
probability to be in the desired initial pure state, along with
probabilities to be in a variety of other undesired states; thus, at
the end of the computation, the answer cannot be read out with high
probability in a single measurement because the signal is buried in
noise produced by the undesirable states.  To extract the signal
requires a number of measurements, made on copies of the physical
system or on repetitions of the computation.  Either of these amounts
to an additional physical resource.  The way this appears in
mixed-state quantum information processors is that the signal is
encoded in an expectation value that can be determined with good
accuracy only from many measurements.

An example is provided by the present method for implementing quantum
information processing in liquid-state NMR. \ The processing elements
are the active molecules in the liquid sample, each of which has $N$
active nuclear spins.  The initialization procedure takes the $N$
nuclear spins from a state of thermal equilibrium, with polarization
$\alpha\sim2\times10^{-5}$, to a so-called {\em pseudopure state\/}
[51,\hspace{2.5pt}52], which has density operator
$\rho=(1-\varepsilon)I/2^N+\varepsilon|\psi\rangle\langle\psi|$. This
density operator is a mixture of the unpolarized, maximally mixed
state of the spins, $I/2^N$, and the desired initial pure state,
$|\psi\rangle\langle\psi|$.  The mixing parameter $\varepsilon$
determines the size of the signal produced by the desired state; a
consequence of pseudopure-state synthesis is that the mixing
parameter decreases exponentially with the number of qubits, i.e.,
$\varepsilon=\alpha N/2^N$.

The exponential signal decrease is an in-principle problem for any
information processing based on pseudopure-state synthesis [54].  To
extract the signal from the random noise produced by the unpolarized
piece of the density operator requires a number of copies or
repetitions that scales as $1/\varepsilon^2=2^{2N}/\alpha^2N^2$, thus
giving rise to an exponential resource demand.  Even given the
macroscopic number ($\sim 10^{20}$) of molecules in an NMR solution,
each of which acts as an independent processor, liquid-state NMR is
limited to about 20 qubits with the initial polarizations presently
available.

Schulman and Vazirani [55] have outlined a method for distilling pure
qubits from the weakly polarized nuclear spins in a liquid NMR
sample.  This method is algorithmic, using operations that can be
implemented in NMR [56].  Though it is highly impractical, requiring
$\sim1/\alpha^2$ initial qubits for each distilled pure qubit, it
does not make an exponential resource demand.  From our perspective,
however, this method is not an example of mixed-state quantum
information processing.  Rather it is a different initialization
procedure, which cools a small subset of the qubits to zero
temperature, using the remaining qubits as a heat reservoir, thus
yielding an initial pure state to which our previous analysis
applies.

We do not have a general analysis of the physical-resource demands
posed by using mixed states for quantum information processing.  We
suspect, however, that computational protocols based on the use of
highly mixed states suffer generally from a demand for an exponential
number of repetitions or copies similar to that for pseudopure-state
synthesis in liquid-state NMR.  This hunch is supported by work [57]
that suggests that supplementing a set of pure qubits with a supply
of maximally mixed qubits provides almost no additional computational
power beyond that in the pure qubits.  These considerations make it
unlikely that systems in highly mixed states can be scalable quantum
computers, but this does not mean that they are equivalent to
classical computers.  They seem to lie somewhere between classical
computers and full-scale quantum computers, since there are special
problems [58,\hspace{2.5pt}59] for which no efficient classical algorithm is known,
but which can be done efficiently using highly mixed states---without
the need for an exponential number of copies or repetitions.

\subsection{Entanglement}
\label{sec:entanglement}

Entanglement is a distinctive feature of quantum mechanics.  It is
clearly a resource for such quantum information protocols as
teleportation, yet its role in quantum computation remains unclear.
Some claim it is the property that powers quantum computation
[47,\hspace{2.5pt}60], while others downplay its significance [61,\hspace{2.5pt}62].  The
situation has been clarified considerably by the recent work of Jozsa
and Linden [63], who showed that for a qubit quantum computer---the
extension to qudits is probably straightforward---entanglement among
all the qubits is a prerequisite for an exponential speed-up over a
classical computation.  The Jozsa-Linden proof proceeds by showing
that if entanglement extends only to some fixed number of qubits,
independent of problem size, the computation can be simulated
efficiently on a classical computer. Jozsa and Linden were careful to
point out that although entanglement among all qubits is necessary
for exponential speed-up, it is not sufficient: as shown by Gottesman
and Knill [2], there are sequences of quantum gates that can be
simulated efficiently even though they entangle all qubits.

The Jozsa-Linden argument assumes a strictly scalable tensor-product
structure.  The global entanglement that accompanies exponential
speed-up is a consequence of assuming this tensor-product structure
and an initial pure state.  This does not necessarily imply that
entanglement is the key resource for quantum computation.  Consider a
computation with an exponential speed-up on a qubit quantum computer.
Mapped onto a unary machine, the same computation produces {\em no\/}
entanglement.  Whether run on the unary computer or the qubit
computer, the computation accesses arbitrary states---i.e., arbitrary
superpositions---in the computer's Hilbert space and has no efficient
description in the realistic language of classical computation.
Hilbert spaces are fungible!  Entanglement is not an inherent feature
of quantum computation, but rather a result of running the
computation on a quantum computer with a tensor-product structure;
for such a computer, arbitrary superpositions lead to entanglement
among all the parts, because the states without such entanglement
occupy only a tiny corner of Hilbert space [47,\hspace{2.5pt}60].  On a physically
unary computer, the same arbitrary superpositions have no
entanglement.

We conclude that the global entanglement in a quantum computation is
a consequence of the need to save resources, which is what dictates a
strictly scalable tensor-product structure to start with.  We suggest
that entanglement, instead of being the power behind quantum
computation, might be a measure of the computer's ability to
economize on physical resources.  This surmise, based on our
consideration of pure-state quantum computation, is supported by what
is known about mixed-state quantum computation in liquid-state NMR. \
The argument that entanglement follows from accessing arbitrary
states in a system with a tensor-product structure doesn't work for
mixed states [63].  Indeed, with present polarizations, the states
accessed in NMR are known to be unentangled up to about 13 qubits
[64,\hspace{2.5pt}65] and, for bigger numbers of qubits, are likely to be far less
entangled than in a corresponding pure-state quantum computer.  This
paucity of entanglement, we suggest, is a signal of the resource
problem in NMR, i.e., the need, discussed above, for exponentially
many molecules in NMR.

To investigate this idea further, one would like to quantify the
amount of global entanglement produced in a quantum computation
carried out in systems ranging from nonscalable to strictly scalable
and including both pure-state and mixed-state realizations.  This is
a daunting task since there is presently no suitable measure of
global entanglement in multipartite quantum systems, even for pure
states.  Indeed, multipartite entanglement cannot be summed up by any
single measure and whether there is a measure or measures tied to the
issue of scalability is far from clear.

\section{Conclusion}
\label{sec:conclusion}

Our contention in this paper is that the fundamental requirement for
a scalable quantum computer is set by the need to economize on
physical resources in providing the primary resource of Hilbert-space
dimension.  To avoid an exponential demand for physical resources,
the number of degrees of freedom---or, for quantum fields, the number
of particles or the number of field modes, depending on which (or
both) acts as effective degrees of freedom---must grow quasilinearly
with the equivalent number of qubits.  These requirements mean that a
scalable quantum computer must have a robust tensor-product
structure.  Systems without such a tensor-product structure are not
suitable for scalable quantum computation.

Physical systems that don't scale properly, such as liquid-state NMR,
Rydberg atoms, or molecular magnets, are still worth studying for a
variety of reasons. First and foremost, they embody fundamental
physical questions that are worth investigating in their own right,
regardless of their relevance to quantum information science. Second,
they can be used to develop new technologies for control, readout,
and error correction in quantum systems.  These new technologies
might have applications to quantum-information-processing jobs
outside quantum computation, and they might be transferable to
scalable quantum computers.  Finally, the scalability criteria
formulated in this paper are asymptotic requirements.  They are
useful for assessing the physical resources required for a
general-purpose quantum computer to do problems of increasing size.
Yet even for this purpose, they are imperfect tools, because no
computer is expected to do problems of arbitrary size.  Nonscalable
systems might be able to provide sufficient Hilbert-space dimension
for special-purpose quantum computations that need only a limited
number of qubits, such as simulation of other quantum systems [66].

Hilbert space is essential for quantum computation.  Yet it is an odd
sort of thing to need.  It is not a physical object, but rather a
mathematical abstraction in which we describe physical objects
[67,\hspace{2.5pt}68].  A Hilbert space gets a physical interpretation---a
connection to the external world---only through the physical system
that we describe in that Hilbert space.  The connection is made
through privileged observables---the generators of space-time
symmetries, e.g., position, momentum, angular momentum, and
energy---which determine a set of physical degrees of freedom for the
system.  This connection made, we can determine how the physical
resources, measured in terms of phase-space actions constructed from
the privileged observables, must grow in order to provide the
Hilbert-space dimension needed for a quantum computation.

Our degrees-of-freedom analysis can be applied to the physical
resources required by a classical computer.  Generally the subsystems
in a classical computer consist of many physical degrees of freedom.
If each distinguishable configuration of a subsystem occupies a fixed
phase-space volume $V_0$, then our analysis shows that the physical
resources required by the classical computer grow exponentially
unless the number of subsystems grows quasipolynomially with problem
size.  But the scale $V_0$ in the classical analysis is not
fundamental, instead being set by noise and the resolution of
measuring devices.  This makes the classical analysis of resource
requirements dependent on other features of a classical computer.
The difference for a quantum computer is that Planck's constant sets
a fundamental scale, which makes the resource requirements presented
here prerequisites for scalable quantum computation, prior to the
other necessary requirements for a quantum computer's operation.

\vspace{12pt}
{\bf Acknowledgments.} This work was partly supported by the National
Security Agency (NSA) and the Advanced Research and Development
Activity (ARDA) under Army Research Office (ARO) Contract
No.~DAAD19-01-1-0648 and by the Office of Naval Research under
Contract Nos.~N00014-00-1-01578 and N00014-99-1-0247.  This work grew
in large part out of discussions at the Institute Theoretical Physics
of the University of California, Santa Barbara, where the authors
were in residence during the fall of 2001.  The authors received
support from the ITP's National Science Foundation Contract
No.~PHY99-07949.

\vspace{24pt}
\noindent
{\Large\bf References}

\begin{enumerate}

\item
{\em Quantum Information Science: An Emerging Field of
Interdisciplinary Research and \linebreak[4]\hbox{Education} in
Science and Engineering}, Report of NSF Workshop, October 28--29,
1999, \hbox{Arlington}, VA, {\tt
http://www.nsf.gov/cgi-bin/getpub?nsf00101}.

\item
M.~A. Nielsen and I.~L.~Chuang, {\em Quantum Computation and Quantum
Information\/} \linebreak[4](\hbox{Cambridge} University Press,
\hbox{Cambridge}, England, 2000).

\item
D.~P.~DiVincenzo, {\em Fortschr.\ Phys.}\ {\bf 48}, 771 (2000).

\item
R.~Raussendorf and H.~J.~Briegel, {\em Phys.\ Rev.\ Lett.}\ {\bf 86},
5188 (2001).

\item
R~Raussendorf and H.~J.~Briegel, ``Computational model underlying the
one-way quantum computer,'' unpublished, {\tt arXiv.org e-print
quant\discretionary{-}{}{-}ph/\discretionary{}{}{}0108067}.

\item
M.~A.~Nielsen, ``Universal quantum computation using only projective
measurement, quantum memory, and the preparation of the $|0\rangle$
state,'' unpublished, {\tt arXiv.org e-print
quant\discretionary{-}{}{-}ph/\discretionary{}{}{}0108020}.

\item
P.~W.~Shor, {\em Phys.\ Rev.~A\/} {\bf 52}, R2493 (1995).

\item
A.~Steane, {\em Proc.\ R.\ Soc.\ London~A\/} {\bf 452}, 2551
(1996).

\item
E.~Knill and R.~Laflamme, {\em Phys.\ Rev.~A\/} {\bf 55}, 900 (1997).

\item
D.~Aharonov and M.~Ben-Or, in {\em Proceedings of the 29th Annual ACM
Symposium on \hbox{Theory} of Computing\/} (ACM Press, New York, 1997),
p.~176; see also {\tt arXiv.org e-print
quant\discretionary{-}{}{-}ph/\discretionary{}{}{}9906129}.

\item
J.~Preskill, {\em Proc.\ R.\ Soc.\ London~A\/} {\bf 454}, 385 (1998).

\item
E.~Knill, R.~Laflamme, and W.~H.~Zurek, {\em Science\/} {\bf 279}, 342
(1998).

\item
E.~Knill, R.~Laflamme, and W.~H.~Zurek, {\em Proc.\ R.\ Soc.\ London~A\/}
{\bf 454}, 365 (1998).

\item
N.~Gisin, G.~Ribordy, W.~Tittel, and H.~Zbinden, {\em Rev.\ Mod.\ Phys.}\
{\bf 74}, 145 (2002).

\item
R.~Cleve, in {\em Quantum Computation and Quantum Information
\hbox{Theory}}, edited by \linebreak[4]C.~\hbox{Macchiavello}, G.~M.
Palma, and A.~Zeilinger (World Scientific, Singapore, 2000), p.~103.

\item
P.~Zanardi, {\em Phys.\ Rev.\ Lett.}\ {\bf 87}, 077901 (2001).

\item
J.~Ahn, C.~Weinacht, and P.~H. Bucksbaum, {\em Science\/} {\bf 287}, 463
(2000).

\item
R.~Zadoyan, D.~Kohen, D.~A. Lidar, and V.~A. Apkarian, {\em Chem.\ Phys.}\
{\bf 266}, 323 (2001); V.~V. Lozovoy and M.~Dantus, {\em Chem.\ Phys.\
Lett.}\ {\bf 351}, 213 (2002).  The authors of both these papers point
out that the Hilbert space of a single molecule is not suitable for
scalable quantum computation.

\item
M.~N. Leuenberger and D.~Loss, {\em Nature\/} {\bf 410}, 789 (2001).

\item
A.~Peres, {\em Quantum \hbox{Theory}: Concepts and Methods\/} (Kluwer,
Dordrecht, 1993), p.~373.

\item
Since in this paper we are interested in comparing how
different systems use physical resources, we use the term {\em
quantum computer\/} for any physical system that has the required
Hilbert-space dimension, and we reserve the term {\em scalable
quantum computer\/} for systems that can provide the required
Hilbert-space dimension efficiently.

\item
L.D. Landau and E.M. Lifshitz, {\em Quantum Mechanics:
Non-Relativistic \hbox{Theory}}, 3rd Ed.\ (\hbox{Oxford} University Press,
\hbox{Oxford}, 1998), p.~172.

\item
We follow the computer-science convention of referring to any
superpolynomial growth as exponential.

\item
We use base-2 logarithms throughout.

\item
D.~A.~Meyer, {\em Science\/} {\bf 289}, 1431a (2000).

\item
P.~G.~Kwiat and R.~J.~Hughes, {\em Science\/} {\bf 289}, 1431a (2000).

\item
M.~W.~Noel and C.~R.~Stroud, Jr., {\em Phys.\ Rev.\ Lett.}\ {\bf 77}, 1913
(1996).

\item
H.~Rabitz, R.~de Vivie-Riedle, M.~Motzkus, and K.~Kompa,
{\em Science\/} {\bf 288}, 824 (2000).

\item
D.~Kielpinski, C.~Monroe, and D.~J. Wineland, {\em Nature\/} {\bf 417},
709 (2002).

\item
P.T. Cochrane and G.~J. Milburn, {\em Phys.\ Rev.~A\/} {\bf 64}, 062312
(2001).

\item
N.~J. Cerf, C.~Adami, and P.~G. Kwiat, {\em Phys.\ Rev.~A\/} {\bf 57},
R1477 (1998).

\item
P.~G. Kwiat, J.~R. Mitchell, P.~D.~D. Schwindt, and A.~G. White,
{\em J.~Mod.\ Opt.}\ {\bf 47}, 257 (2000).

\item
J.~F. Clauser and J.~P.~Dowling, {\em Phys.\ Rev.~A\/} {\bf 53}, 4587
(1996).

\item
The exponential growth of physical resources required for readout in
a classical wave computer and its relation to a unary quantum
computer using a single photon has been noted by S.~Wallentowitz,
I.~A. Walmsley, and J.~H. Eberly, ``How big is a quantum computer?''
unpublished, {\tt arXiv.org e-print
quant\discretionary{-}{}{-}ph/\discretionary{}{}{}0009069}.

\item
N.~Bhattacharya, H.~B. van Linden van den Heuvell, and R.~J.~C. Spreeuw,
{\em Phys.\ Rev.\ Lett.}\ {\bf 88}, 137901 (2002).

\item
Grover's algorithm does not fit into our resource discussion because
it does not provide an exponential speed-up relative to classical
algorithms.  Nonetheless, the experiment in Ref.~35 illustrates the
resource demands that come with using classical waves.

\item
G.~J. Milburn, {\em Phys.\ Rev.\ Lett.}\ {\bf 62}, 2124 (1989).

\item
E.~Knill, R.~Laflamme, and G.~J. Milburn, {\em Nature\/} {\bf 409},
46 (2001).

\item
S.~D. Bartlett, B.~C. Sanders, S.~L. Braunstein, and K.~Nemoto,
{\em Phys.\ Rev.\ Lett.}\ {\bf 88}, 097904 (2002).

\item
S.~D. Bartlett and B.~C. Sanders, ``Efficient classical simulation of
optical quantum circuits,'' unpublished, {\tt arXiv.org e-print
quant\discretionary{-}{}{-}ph/\discretionary{}{}{}0204065}.

\item
S.~B.~Bravyi and A.~Yu.~Kitaev, ``Fermionic quantum computation,''
unpublished, {\tt arXiv.org e-print
quant\discretionary{-}{}{-}ph/\discretionary{}{}{}0003137}.

\item
B.~M.~Terhal and D.~P.~DiVincenzo, {\em Phys.\ Rev.~A\/} {\bf 65},
032325 (2002).

\item
E. Knill, ``Fermionic linear optics and matchgates,'' unpublished,
{\tt arXiv.org e-print \linebreak[4]
quant\discretionary{-}{}{-}ph/\discretionary{}{}{}0108033}.

\item
A.~Barenco, C.~H. Bennett, R.~Cleve, D.~P. DiVincenzo, N.~Margolus,
P.~Shor, T.~Sleator, J.~A. Smolin, and H.~Weinfurter, {\em Phys.\
Rev.~A\/} {\bf 52}, 3457 (1995).

\item
J.~L.~Dodd, M.~A.~Nielsen, M.~J.~Bremner, and R.~T.~Thew, {\em Phys.\
Rev.~A\/} {\bf 65}, 040301(R) (2002).

\item
M.~A.~Nielsen, M.~J.~Bremner, J.~L.~Dodd, A.~M.~Childs, and
C.~M.~Dawson, ``Universal simulation of Hamiltonian dynamics for
qudits,'' unpublished, {\tt arXiv.org e-print
quant\discretionary{-}{}{-}ph/\discretionary{}{}{}0109064}.

\item
A.~Ekert and R.~Jozsa, {\em Phil.\ Trans.\ R.\ Soc.\ London~A\/}
{\bf 356}, 1769 (1998).

\item
W.~H.~Zurek, {\em Phys.\ Today\/} {\bf 44}(10), 36 (1991).

\item
D.~Giulini, E.~Joos, C.~Kiefer, J.~Kupsch, I.-O.~Stamatescu, and
H.~D. Zeh, {\em Decoherence and the Appearance of a Classical World
in Quantum \hbox{Theory}\/} (Springer, Berlin, 1996).

\item
J.~R. Anglin, J.~P. Paz, and W.~H. Zurek, {\em Phys.\ Rev.~A\/} {\bf
55}, 4041 (1997).

\item
D.~G. Cory {\em et al.}, {\em Fortschr.\ Phys.}\ {\bf 48}, 875 (2000).

\item
J.~A. Jones, {\em Fortschr.\ Phys.}\ {\bf 48}, 909 (2000).

\item
L.~M.~K. Vandersypen, M.~Steffen, G.~Breyta, C.~S. Yannoni, M.~H.
Sherwood, and I.~L. Chuang, {\em Nature\/} {\bf 414}, 883 (2001).

\item
W.~S.~Warren, {\em Science\/} {\bf 277}, 1688 (1997).

\item
L.~J.~Schulman and U.~V. Vazirani, in {\em Proceedings of the 31st
Annual ACM Symposium on \hbox{Theory} of Computing\/} (ACM Press, New
York, 1999), p.~322; see also {\tt arXiv.org e-print
quant\discretionary{-}{}{-}ph/\discretionary{}{}{}9804060}.

\item
D.~E. Chang, L.~M.~K. Vandersypen, and M.~Steffen, {\em Chem.\ Phys.\
Lett.}\ {\bf 338}, 337 (2001).

\item
A.~Ambainis, L.~J.~Schulman, and U.~V.~Vazirani, in {\em Proceedings
of the 32nd Annual ACM Symposium on \hbox{Theory} of Computing\/}
(ACM Press, New York, 2000), p.~697; see also \linebreak[4]{\tt
arXiv.org e-print
quant\discretionary{-}{}{-}ph/\discretionary{}{}{}0003136}.

\item
E.~Knill and R.~Laflamme, {\em Phys.\ Rev.\ Lett.}\ {\bf 81}, 5672 (1998).

\item
R.~Laflamme, D.~G.~Cory, C.~Negrevergne, and L.~Viola, {\em Quant.\ Inf.\
Comp.}\ {\bf 2}, 166 (2002).

\item
R.~Jozsa, in {\em The Geometric Universe: Science, Geometry, and the
Work of Roger Penrose}, edited by S.~A. Huggett, L.~J. Mason, K.~P.
Tod, S.~T. Tsou, and N.~M.~J. Woodhouse (\hbox{Oxford} University Press,
\hbox{Oxford}, England, 1998), p.~369.

\item
S.~Lloyd, {\em Phys.\ Rev.~A\/} {\bf 61}, 010301(R) (1999).

\item
P.~Knight, {\em Science\/} {\bf 287}, 441 (2000).

\item
R.~Jozsa and N.~Linden, ``On the role of entanglement in quantum
computational speed-up,'' unpublished, {\tt arXiv.org e-print
quant\discretionary{-}{}{-}ph/\discretionary{}{}{}0201143}.

\item
S.~L. Braunstein, C.~M. Caves, R.~Jozsa, N.~Linden, S.~Popescu, and
R.~Schack, {\em Phys.\ Rev.\ Lett.}\ {\bf 83}, 1054 (1999).

\item
N.~C. Menicucci and C.~M.~Caves, {\em Phys.\ Rev.\ Lett.}\ {\bf 88},
167901 (2002).

\item
S.~Lloyd, {\em Science\/} {\bf 273}, 1073 (1996).

\item
C.~M.~Caves, C.~A.~Fuchs, and R.~Schack, {\em Phys.\ Rev.~A\/} {\bf 65},
022305 (2002).

\item
C.~A. Fuchs has promoted the notion that Hilbert-space dimension is a
``characteristic property'' of a quantum system, {\tt arXiv.org
e-print quant\discretionary{-}{}{-}ph/\discretionary{}{}{}0204146},
or perhaps an ``element of reality,'' {\tt arXiv.org e-print
quant\discretionary{-}{}{-}ph/\discretionary{}{}{}0205039}, which
might be the realistic core of quantum theory.

\end{enumerate}

\end{document}